\documentstyle[epsfig]{mn}
\begin{document}

\LARGE
\normalsize

\title[Magellanic clouds radio survey]
{A radio survey of supersoft, persistent and transient
X-ray sources in the Magellanic Clouds}
\author[R.~P.~Fender et al.]
{R. P. Fender$^{1,2}$
K. Southwell$^3$\thanks{Present address : Nature, 4 Crinan Street,
London N1 9XW}
A. K. Tzioumis$^4$\\
$^1$Astronomy Centre, University of Sussex, Falmer, Brighton BN1 9QH, UK\\ 
$^2$Astronomical Institute `Anton Pannekoek', University of Amsterdam,
and Centre for High Energy Astrophysics, Kruislaan 403, \\
1098 SJ Amsterdam, The Netherlands \\
$^3$Department of Astrophysics, Nuclear Physics Building, Keble Road,
Oxford OX1 3RH\\
$^4$Australia Telescope National Facility, CSIRO, PO Box 75, Epping 2121, NSW,
Australia\\}

\maketitle

\begin{abstract}

We present a radio survey of X-ray sources in the Large and Small
Magellanic clouds with the Australia Telescope Compact Array at 6.3
and 3.5 cm.  Specifically, we have observed the fields of five LMC
and two SMC supersoft X-ray sources, the X-ray binaries LMC X-1, X-2,
X-3 \& X-4, the X-ray transient Nova SMC 1992, and the soft gamma-ray
repeater SGR 0525-66.  None of the targets are detected as point
sources at their catalogued positions. In particular, the proposed
supersoft jet source RXJ 0513-69 is not detected, placing constraints
on its radio luminosity compared to Galactic jet sources.  Limits on
emission from the black hole candidate systems LMC X-1 and X-3 are
consistent with the radio behaviour of persistent Galactic black hole
X-ray binaries, and a previous possible radio detection of LMC X-1 is
found to almost certainly be due to nearby field sources. The SNR
N49 in the field of SGR 0525-66 is mapped at higher resolution than
previously, but there is still no evidence for any enhanced emission
or disruption of the SNR at the location of the X-ray source.

\end{abstract}


\section{Introduction}

Radio synchrotron and X-ray emission, though at opposite ends of the
electromagnetic spectrum, are both tracers of high-energy phenomena in
astrophysical sources. Thermal X-ray emission clearly demonstrates the
existence of material at extremely high temperatures, while radio
synchrotron emission originates in the spiralling of highly
relativistic electrons around magnetic field lines. It has become
increasingly apparent that the behaviour of sources in one energy
regime may be correlated with that in another, although the exact
mechanism is often unclear. In particular, radio and X-ray emission
from compact Galactic X-ray sources are often related. While radio
emission from such sources, typically at distances between 1 - 10 kpc,
is now relatively routinely detected and monitored, it has yet to be
detected from a binary in an extragalactic system. With this in mind,
we have searched for radio emission from some of the most powerful
X-ray emitters in the nearest external galaxies, the Large and Small
Magellanic clouds, at $\sim 55$ and $\sim 60$ kpc respectively.

Targets for our survey included most of the Magellanic cloud
{\em supersoft X-ray sources}, several bright LMC X-ray binaries
and other transient systems.

\subsection{Supersoft sources}

The prototypical supersoft X-ray sources, CAL~83 and CAL~87, were
first detected in the Large Magellanic Cloud in 1979-1980 with the
{\it Einstein} X-ray Observatory (Long, Helfand \& Grabelsky, 1981),
although later {\it ROSAT} observations have considerably enlarged the
group (Tr\"{u}mper et al.\ 1991). The defining characteristics of the
supersoft sources are their extremely low X-ray
energies and high bolometric luminosities (typically $L_{\rm bol}
\sim 10^{38}$~erg\,s$^{-1}$ and T$_{\rm bb} \sim$~tens of eV, where
T$_{\rm bb}$ is the blackbody temperature).

Progress in determining the exact nature of these systems has been
hindered by the fact that they are undetectable in the Galactic plane,
due to the high level of soft X-ray absorption (e.g. van den Heuvel
et al.\ 1992). Most currently known systems are therefore optically
faint extragalactic objects, predominantly in the Magellanic clouds
and M31 (see e.g. Kahabka \& Tr\"umper 1996 for a review). Although
the term ``supersoft source'' has been previously applied to a range
of objects such as planetary nebula nuclei (Wang 1991) and PG~1159
stars (Cowley et al.\ 1995), we shall consider here only those objects
exhibiting the characteristics of X-ray binaries (e.g. Crampton et
al.\ 1987; Smale et al.\ 1988; Pakull et al.\ 1988; Cowley et al.\
1990; Pakull et al.\ 1993).

The model for the SSSs which has gained predominanceis that of an
accreting white dwarf in a binary system which is undergoing steady
nuclear burning on its surface (van den Heuvel et al.\ 1992). However,
it should be noted that models of black hole (Cowley et al.\ 1990;
Crampton et al.\ 1996) and neutron star accretors (Greiner, Hasinger
\& Kahabka 1991; Kylafis \& Xilouris 1993) also exist. Indeed, one of
the systems considered in this paper, RXJ~0059-71, almost certainly
contains a neutron star, although, with 2.7~s pulsations (Hughes 1994)
and a Be-type secondary star (Southwell \& Charles 1996), it is not
strictly an archetypal source.

Of the other supersoft objects considered here, RXJ~0513-69 is unique
in being the only SSS to exhibit optical jets (Pakull 1994, private
communication; Cowley et al.\ 1996).  The source is an X-ray transient
which was discovered in outburst during the {\it ROSAT} All Sky Survey
(Schaeidt, Hasinger \& Tr\"{u}mper 1993). The optical spectrum (Pakull
et al.\ 1993; Cowley et al.\ 1993; Crampton et al.\ 1996; Southwell et
al.\ 1996) is similar to that of CAL~83, and the two sources have
comparable optical magnitudes (V $\sim 16-17$~mag). However, only
RXJ~0513-69 exhibits Doppler-shifted components of He{\sc ii}~4686 and
H$\beta$, with velocities characteristic of the escape speed of a
white dwarf (Southwell et al.\ 1996), implying the presence of a
highly-collimated outflow. However, despite the drawing of analogies
with SS433, neither this source nor any other supersoft X-ray binary
have ever been detected at radio wavelengths.

One of the LMC sources considered here, RX~J0550-71, does not yet have
an optical counterpart, hence it should be noted that the nature of
this object is particularly uncertain.

\begin{table*}
\begin{minipage}{120mm}
\caption{Radio survey of Magellanic Cloud X-ray sources with the 
Australia Telescope compact array. All upper
limits are 3$\sigma$.}
\begin{tabular}{cccccc}
 Source & Object &\multicolumn{4}{c}{Point-source flux density (mJy)} \\
        & type &  3.5 cm & 6.3 cm & 12.7 cm & 21.7 cm\\
\hline
RXJ 0513-69 & LMC supersoft source & $<0.09$ & $<0.09$ & $<0.15$ & $<0.18$ \\
RXJ 0528-69 & LMC supersoft source & $<0.09$ & $<0.09$ & &\\
CAL 83 & LNC supersoft source & $<0.12$ & $<0.12$ & &\\
CAL 87 & LMC supersoft source & $<0.12$ & $<0.12$ & & \\
RXJ 0550-71 & LMC supersoft source & $<0.15$ & $<0.15$ & & \\
LMC X-1 & BHC X-ray binary & $<1.5$ & $<1.5$ & & \\
LMC X-2 & X-ray binary & $<0.15$ & $<0.15$ & &\\
LMC X-3 & BHC X-ray binary & $<0.12$ & $<0.18$ & & \\
LMC X-4 & X-ray pulsar X-ray binary & $<0.15$ & $<0.18$ & &\\
SGR 0525-66 & Soft $\gamma$-ray repeater & $<0.3$ & $<0.6$ & &\\
\hline
1E 0035-72 & SMC supersoft source & $<0.12$ & $<0.12$ &  & \\
RXJ 0059-71 & SMC supersoft source & $<0.12$ & $<0.15$ & & \\
Nova SMC 1992 & SMC X-ray transient & $<0.18$ & $<0.15$ & &\\
\hline
\end{tabular}
\end{minipage}
\end{table*}

\subsection{Radio emission from X-ray binaries}

Radio emission has been detected from approximately 20\% of Galactic
X-ray binary systems, comprising a neutron star or black hole
accreting matter from a more normal companion (e.g. Hjellming \& Han
1995). In several cases the emission has been resolved by
high-resolution observations into jet-like structures reminiscent of
outflows from AGN, and relativistic or near-relativistic velocities
inferred (e.g. Fender, Bell~Burnell \& Waltman 1997 and references
therein). It now seems that black hole systems are
particularly likely sources of radio emission, whether transient
or persistent -- in either case the characteristics of the radio
emission are mirrored in those of the X-ray emission.

Persistent black hole systems in quiescence appear to have centimetric
radio luminosities which agree within a factor two with each
other. For example, observations of Cygnus X-1, GX 339-4,
1E1740.7-2942 and GRS 1758-258 are all consistent with a centimetric
flux density of $\sim 10$ mJy at 3 kpc. More transient systems such as
Cygnus X-3, GRS 1915+105 and X-ray `novae' such as V404 Cyg, show much
more dramatic variability, with measured flux densities from $<1$ to
$>20 000$ mJy.

LMC X-1 and LMC X-3, two very luminous black hole candidates may be
included in the class of persistent X-ray sources. At a distance of 55
kpc, we would only expect to observe a flux density of a few tens of
$\mu$Jy, by analogy with their Galactic cousins. However, lack of
knowledge of their true nature and the chance of catching a rare
flaring state makes the observations worthwhile. In the case of LMC
X-1 this is particularly so, as Spencer et al. (1997) report the
detection of a significant ($\sim 80$ mJy) flux from the field of
this source. 

LMC X-4 is an X-ray pulsar system, containing a highly magnetised
accreting neutron star. Such systems in our Galaxy are found not to be
radio-emitters (Fender et al. 1997). LMC X-2 is a low-mass X-ray
binary system thought to contain a neutron star. While we considered
both these sources far less likely to be detected than LMC X-1 or LMC
X-3, they were included in the survey for completeness.

Nova SMC 1992 (Clark, Remillard \& Woo 1996) was discovered in
archival ROSAT observations of 1992 Oct 1-2 as an extremely bright
transient X-ray source. Clark et al. (1996) proposed a nearby 14th
magnitude blue star as the optical counterpart, and suggested the
source may be the first high-mass black hole X-ray nova
detected. While previous X-ray `novae' have often been very bright
radio sources (see above), they are also generally associated with
low-mass companion stars and the nature of this system remains
uncertain.

\subsection{SGR 0525-66}

SGR 0525-66 is one of a group of only three confirmed (Smith 1997) and
one possible additional (Hurley et al. 1997) soft $\gamma$-ray
repeaters. These are sources of repeated bursts of low-energy
$\gamma$-rays, possibly associated with neutron stars. SGR 0525-66
is coincident in the sky with the SNR N49, and a soft X-ray counterpart
has been found (Rothschild, Kulkarni \& Lingenfelter 1994). No radio,
optical or infrared counterpart has been identified however (Dickel
et al. 1995), although Rothschild et al. (1994) predicted a compact
plerion-like radio nebula to be associated with the source. Any
physical link with N49 has yet to be established.

\section{Observations}

Observations of the target sources were made between 1997 June 30 and
July 13 with the Australia Telescope Compact Array. The array was in
the 6D configuration, with baselines ranging between 153 and 6000
m. All observations were made simultaneously at 6.3 and 3.5 cm, except
for the RXJ 0513-69 field, which was also observed at 20 \& 13 cm, on
July 13.  Primary flux calibration was achieved in all cases using PKS
1934-638; phase calibration was achieved using calibrators 0515-674
and 0252-712 for the LMC and SMC fields respectively.  Observations
typically involved 5 min on a nearby phase calibrator followed by 25
min on-target, repeated over a 12-hr run (though less time was spent
on LMC X-1, X-2, X-3 and X-4). Data reduction was performed using the
MIRIAD software package running on the Sussex University STARLINK
node.

No radio point sources were detected at any wavelength at the
catalogued locations of the target sources. In most cases a
noise level of 50 $\mu$Jy or so was achieved, making the 
3$\sigma$ upper limits very stringent indeed. The noise levels
for LMC X-1 and SGR 0525-66 are considerably worse, due to 
their locations in radio-bright regions (see below). Table 1
summarizes the results and lists 3$\sigma$ upper limits to
point-source emission at the target locations.

\begin{figure}
\leavevmode\epsfig{file=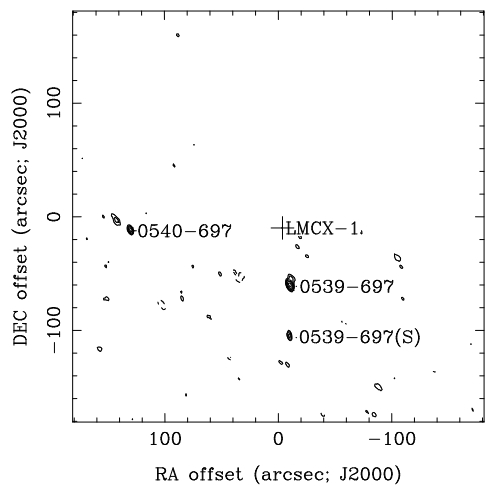,width=7.5cm,angle=270,clip,
bbllx=10,bblly=50,bburx=630,bbury=750}
\caption{6.3 cm radio map centred on optical location of the black
hole candidate LMC X-1. The nearby sources 0540-697 and 0539-697
(Wright et al. 1994), plus the previously unidentified source,
labelled 0539-697(S), are probably the cause of a previous report of
radio emission from this region. The apparent alignment of
0539-697(S), 0539-697, and the optical position of LMC X-1, while
reminiscent of arcmin-scale radio lobes associated with 1E1740.7-2942,
is probably coincidental. Contours are at -3, 3, 6, 12, 24, 48 and 96
times the r.m.s. noise of 0.5 mJy.}
\end{figure}

\section{Discussion}

\subsection{Supersoft sources}

The upper limit to radio emission from RXJ 0513-6951 indicates it is
at least a factor of 50 less luminous at radio wavelengths than the
classical Galactic jet source SS 433, which also shows Doppler shifted
optical emission lines.  However, the jets of SS 433 have a velocity
of 0.26 c, while those of RXJ 0513-6951 are less than 0.01 c, and so
clearly far less energetic.  It may be that the supersoft source
simply does not accelerate electrons to the relativistic energies
required to produce synchrotron emission.

No radio nebulosity, on scales from a few arcsec to a few arcmin, was
detected associated with any of these systems, expected to be strong
sources of ionising UV photons (although note that with a minimum
baseline of 153m this configuration of the array is not sensitive to
structureswith angular scales $\geq$ 1 arcmin).  In particular, the
arcmin-scale optical nebula around CAL83 (Remillard et al. 1995) was
not observed.

\subsection{X-ray binaries}

\begin{figure*}
\leavevmode\epsfig{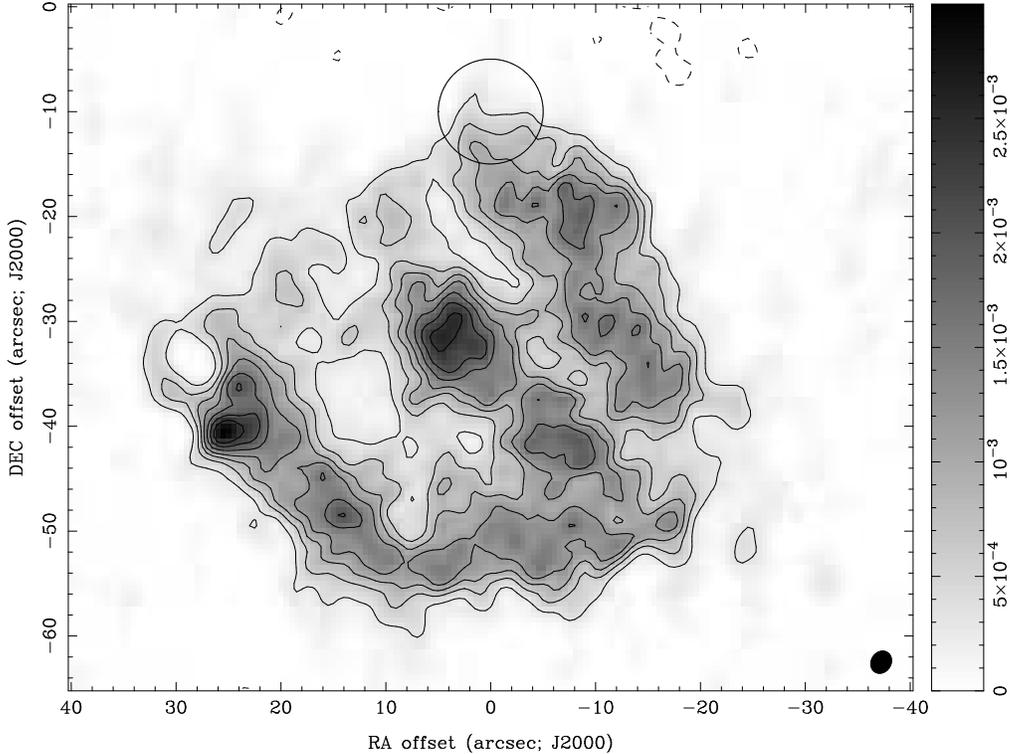}
\caption{6.3 cm radio map of the SNR N49, with the X-ray location of
SGR 0525-66 (Rothschild et al. 1994) indicated as a 5 arcsec radius
circle.  There is no evidence for enhanced radio emission, nor
disruption of the SNR, at the position of SGR 0525-66. Contours are at
-3, 3, 6, 9, 12, 16, 20 and 24 times 0.1 mJy. The synthesised beam is
2.30 $\times$ 1.96 arcsec (p.a. -33$^{\circ}$) and is indicated in the
lower right corner of the map. The map was created with a robust
factor of 0.5 and 5000 CLEAN iterations with a loop gain of 0.1.}
\end{figure*}

\subsubsection{LMC X-1 and LMC X-3}

Neither of these two black hole candidates were detected to levels
consistent with analogy to their Galactic counterparts. At a 
distance of 55 kpc, the persistent Galactic radio-emitting X-ray
binaries would only be observed at a level of $\sim 0.05$ mJy,
at about the noise level for LMC X-3, and well below it for LMC X-1.

The noise level in the LMC X-1 observations is far higher than that
for the other fields observed; this is due to it residing along a very
radio-bright line of sight in the LMC.  The source lies at the edge of
the bright HII region N149, and in addition there are numerous
confusing sources at a level of several tens of mJy or brighter (see
e.g. Chu et al. 1997).  The possible radio detection of this source
noted by Spencer et al. (1997) in their southern hemisphere radio
survey of X-ray binaries is almost certainly due to the two radio
sources 0540-697 and 0539-697, previously catalogued by Wright et
al. (1994), and indicated in Fig 1. Gaussian fits to these two sources
give flux densities of $\sim 36$ and $\sim 105$ mJy respectively at
6.3 cm. We note the presence of an additional uncatalogued radio
source, labelled 0539-697(S) in Fig 1, some 45 arcsec south of
0539-697, for which Gaussian fitting gives a 6.3 cm flux density of
$\sim 30$ mJy. This source, 0539-697, and LMC X-1 are all roughly
aligned, reminiscent of large-scale radio lobes associated with
1E1740.7-2942 (Mirabel et al. 1992), but we feel the alignment is
probably coincidental. These results are consistent with a more
detailed analysis of the southern hemisphere survey data
(R. E. Spencer, private communication).

\subsubsection{LMC X-2 and LMC X-4}

The lack of detection of these two sources is again unsurprising.
In particular, as noted above, we do not expect to observe radio
emission from an X-ray pulsar system whatever the distance.

\subsubsection{Nova SMC 1992}

The lack of detection of this source is not surprising, nearly 
four years after its `X-ray nova' outburst, as most X-ray
transients return to radio quiescence within a year or so of
outburst. 

\subsection{SGR 0525-66}

An image of the field of SGR 0525-66 at 6.3 cm is shown in Fig 2, with
the location of the source at the upper bound of the N49 SNR
indicated. Details of the map derivation are given in the figure
caption.  While having poorer {\em u-v} coverage than the maps of
Dickel et al. (1995), our map is sensitive to higher-resolution
features.  The relatively high 3$\sigma$ point-source upper limit
reflects a high brightness background due to the SNR. In agreement
with Dickel et al. (1995) we find neither enhanced point-source
emission at the location of the SGR, nor any evidence for disruption
of the SNR either at its location or in a path extending back to the
core.

It should be stressed that these observations are sensitive only to
structures on arcsecond scales, and are resolving out much of the
emission from the SNR on arcmin and larger angular
scales. Nevertheless, the prediction of Rothschild et al. (1994) that
SGR 0525-66 should have an associated compact ($\leq 0.3$ arcsec)
synchrotron nebula can be tested. The X-ray point source that they
have detected with ROSAT has an X-ray luminosity of $\sim 10^{36}$ erg
s$^{-1}$. Using the ratio of X-ray to radio luminosities of 1 -- 300
established in e.g. Helfand \& Becker (1987) for synchrotron nebulae,
we would expect a radio luminosity $\geq 10^{33}$ erg s$^{-1}$ if the
origin of the X-ray emission is indeed a nonthermal compact
synchrotron nebula, or `plerion'. However, even at the distance of the
LMC (which we here assume to be 55 kpc), our radio observations place
limits on the radio luminosity of an associated synchrotron nebula of
around $5 \times 10^{31}$ erg s$^{-1}$, significantly below that which
would be expected for a plerion. Hence these radio observations cast
doubt upon the speculation of Rothschild et al. (1994), as did the
earlier observations of Dickel et al. (1995).

\section{Conclusions}

We have surveyed the fields of eight LMC and SMC supersoft
X-ray sources, the X-ray binaries LMC X-1, X-2, X-3 \& X-4, and
the soft $\gamma$-ray repeater SR 0525-66 at radio wavelengths.
We have found no point-source radio emission from any of the
sources.

In particular we find no detectable radio emission from the 
proposed jet source RXJ 0513.9-6951. Neither do we detect
nebulosity, such as that observed optically around CAL83,
associated with any of the supersoft sources.

Limits on emission from the black hole candidate X-ray binaries LMC
X-1 and LMC X-3 are consistent with the radio brightnesses of their
Galactic analogues. We show that a possible previous radio detection
of LMC X-1 was almost certainly due to nearby field sources, and that
due to its location in a radio-bright part of the LMC, this source is
going to be very difficult to ever detect. Limits on radio emission
from the other two X-ray binaries are as expected. 

The SNR N49, which SGR 0525-66 appears to lie on the northern edge of,
shows no enhanced emission at the location of the SGR, nor any
disruption to its structure suggesting association between the two.
We can constrain the radio luminosity of any compact (arcsec-scale)
structure associated with the SGR to be more than an order of
magnitude below that which we might expect if the X-ray source
discussed in Rothschild et al. (1994) were indeed a synchrotron nebula
powered by the SGR.

In summary, we have placed limits on radio emission from a variety 
of X-ray sources in the Magellanic Clouds, finding nore of them
to be anomalously bright by comparison with the Galactic counterparts.
The detection of radio emission from extragalactic X-ray binaries
and related systems is likely to require an increase in the sensitivity of
ground-based arrays or coordinated and fortunate observations during an
outburst.

\section*{acknowledgements}

We are happy to acknowledge assistance with the observations by Vince
McIntyre, and useful discussions with Ralph Spencer.  We also thank
the referee, David Helfand, for useful suggestions which improved the
paper.  The Australia Telescope is funded by the Commonwealth of
Australia for operation as a National Facility managed by CSIRO.

\end{document}